\documentclass[11pt,fleqn]{article}

\usepackage{amsfonts}
\usepackage{amsmath}
\usepackage{amssymb}
\usepackage{mathtools}
\usepackage{mathrsfs}
\usepackage{enumerate}
\usepackage{bbm}
\usepackage{graphicx,epsfig,amsthm,color}
\usepackage{pstricks}
\usepackage{subfig}
\usepackage{graphicx}
\usepackage{units}

\usepackage{caption}
\usepackage{booktabs}
\usepackage{textcomp}
\usepackage{array}
\usepackage{cite}
\usepackage{cancel}

\date{\today}

\newcolumntype{z}[1]{>{\RaggedRight\hspace{0pt}}p{#1}}
\newcolumntype{w}[1]{>{\RaggedRight\hspace{0pt}}p{#1}}
\newcolumntype{v}[1]{>{\Centering\hspace{0pt}}p{#1}}

\def\be{\begin{equation}}
\def\ee{\end{equation}}
\def\bea{\begin{eqnarray}}
\def\eea{\end{eqnarray}}

\def\be{\begin{equation}}
\def\ee{\end{equation}}
\def\bea{\begin{eqnarray}}
\def\eea{\end{eqnarray}}

\def\erp2{{\rm e}^{2\rho}}
\def\erm2{{\rm e}^{-2\rho}}
\def\er4{{\rm e}^{4\rho}}

\def\be{\begin{equation}}
\def\ee{\end{equation}}
\def\bea{\begin{eqnarray}}
\def\eea{\end{eqnarray}}

\def\m0{m_{\nu_{0,i}}}

\def\T0{T_{\nu_0}}

\newcommand{\half}{\frac{1}{2}}

\newcommand{\beqa}{\begin{eqnarray}}
\newcommand{\eeqa}{\end{eqnarray}}
\newcommand{\bpr}{\begin{problem}}
\newcommand{\epr}{\end{problem}}
\newcommand{\bcent}{\begin{center}}
\newcommand{\ecent}{\end{center}}
\newcommand{\bfig}{\begin{figure}}
\newcommand{\efig}{\end{figure}}
\newcommand{\bpc}{\begin{picture}}
\newcommand{\epc}{\end{picture}}

\newcommand{\nnb}{\nonumber}
\newcommand{\reflef}{(\ref}

\newcommand{\nn}{\nonumber}

\renewcommand{\and}{A_{0}^{\nu ,D}(s)}

\newcommand{\bee}{\begin{equation}}

\def\beq{\begin{eqnarray}}
\def\eeq{\end{eqnarray}}

\newcommand{\bright}{\begin{flushright}}
\newcommand{\eright}{\end{flushright}}
\newcommand{\bminip}{\begin{minipage}}
\newcommand{\eminip}{\end{minipage}}

\DeclareCaptionLabelSeparator{mysep}{\hspace{3pt}:\hspace{3pt}}
\DeclareCaptionLabelFormat{mypiccap}{Fig.\hspace{3pt}{#2}}
\DeclareCaptionLabelFormat{mytabcap}{Tab.\hspace{3pt}{#2}}

\begin{document}

\date{}
\title{
\vskip 2cm {\bf\huge Chameleonic equivalence postulate and wave function collapse}\\[0.8cm]}

\author{
{\sc\normalsize Andrea Zanzi\footnote{Email: zanzi@th.physik.uni-bonn.de}\!\!}\\[1cm]
{\normalsize Via Pioppa 261, 44123 Pontegradella (Ferrara) - Italy}\\}
 \maketitle \thispagestyle{empty}
\begin{abstract}
{A chameleonic solution to the cosmological constant problem and the non-equivalence of different conformal frames at the quantum level have been recently suggested [Phys. Rev. D82 (2010) 044006]. In this article we further discuss the theoretical grounds of that model and we are led to a chameleonic equivalence postulate (CEP). Whenever a theory satisfies our CEP (and some other additional conditions), a density-dependence of the mass of matter fields is naturally present. Let us summarize the main results of this paper. 1) The CEP can be considered the microscopic counterpart of the Einstein's Equivalence Principle and, hence, a chameleonic description of quantum gravity is obtained: in our model, (quantum) gravitation is equivalent to a conformal anomaly. 2) To illustrate one of the possible applications of the CEP, we point out a connection between chameleon fields and quantum-mechanical wave function collapse. The collapse is induced by the chameleonic nature of the theory. We discuss the collapse for a Stern-Gerlach experiment and for a diffraction experiment with electrons. More research efforts are necessary to verify whether these ideas are compatible with phenomenological constraints.}
\end{abstract}

\clearpage

\tableofcontents
\newpage

\setcounter{equation}{0}
\section{Introduction}

One of the main problems in modern cosmology is the cosmological constant (CC) one \cite{Weinberg:1988cp} (for a review see \cite{Nobbenhuis:2006yf}). Current observational data are compatible with a meV CC, however, theoretical predictions of the CC are, typically, very different from the meV scale. Let us start considering the Standard Model (SM) of particle Physics. It is common knowledge that the Standard Model (SM) provides an extremely successful description of electroweak and strong interactions up to (roughly) the TeV scale. On the one hand, if we decide to choose the TeV scale as UV cut-off, the theoretical prediction of the vacuum energy (which contributes to the CC) is already much larger than the observed value of the CC (even if the TeV scale is much smaller than the Planck one). Therefore, a naturalness problem must be faced: the meV scale is much smaller than typical particle Physics mass scales. On the other hand, one might think that this is not a big problem, because the idea of neglecting the vacuum energy in the SM is not unusual. After all, the vacuum energy is important in problems with boundaries (for example Casimir energy) or in models including gravitation (because energy gravitates) and the SM does not include gravity. How can we describe gravity? Many theories of gravitation are available today, however, the reference theory is Einstein's General Relativity (GR). GR does not make a prediction of the CC and we are free to set its value to zero as Einstein did (see for example \cite{Witten:2000zk} for a discussion of this point). For this reason, the CC problem is not so acute granted that we remain in the framework of classical gravity (in our example GR) and this remains true even if we include a quantum description of the remaining interactions exploiting the SM. Obtaining a reasonable CC from SM+GR may require a fine-tuning, but the CC problem becomes really acute only when we move to the quantum gravity (QG) regime. In this case, indeed, the CC is not a free parameter anymore and, moreover, we have to understand how the vacuum energy gravitates on cosmological distances. Once this task is accomplished, we have to find a mechanism to keep under control the contribution to the CC of an infinite number of quantum diagrams (including graviton loops). One crucial aspect of this analysis is the predictive nature of our QG theory with respect to the CC. In other words: do we manage to obtain a prediction of the CC from our QG theory? It is common knowledge that our chances to make a prediction in the SM are related to the renormalizability of the theory: QED or QCD are not finite theories, nevertheless they are predictive once the renormalization process is properly taken into account. The situation is similar in QG: the CC problem is closely related to the renormalizability of our QG theory. Needless to say, the CC problem is much more acute than the unpalatable fine-tuning of the SM+GR scenario mentioned above.

In a recent paper \cite{Zanzi:2010rs} we proposed a stringy solution to the CC problem and we pointed out the non-equivalence of different conformal frames at the quantum level. Among the crucial elements of our analysis \cite{Zanzi:2010rs}, we can mention, on the one hand, a splitting of all the fields into a local fluctuating component and a global background one, on the other hand, the peculiar ''chameleonic'' behaviour of the Einstein-frame (E-frame) dilaton $\sigma$. Let us start considering the element we mentioned last. In our model the E-frame dilaton belongs to that group of scalar fields (introduced in the literature in \cite{Khoury:2003aq, Khoury:2003rn, Mota:2003tc}) coupled to matter (including the baryonic one) and with an increasing mass as a function of the matter density of the environment. 
In other words, the physical properties of this field
vary with the matter density of the environment and, therefore, it
has been called chameleon. This is one of the possible ways in which local tests of gravity can be faced. Among the other possibilities already discussed in the literature, we mention: 1) symmetron theories \cite{Hinterbichler:2010es, Olive:2007aj, Pietroni:2005pv} (where fifth-forces are screened through a restoration of symmetry at high densities); 2) Galileon theories \cite{Nicolis:2008in}, where non-canonical kinetic terms reduce the effective coupling to matter. Moreover, the chameleon mechanism can be considered as a stabilization mechanism. Many other stabilization
mechanisms have been studied for the string dilaton in the
literature. In particular, as far as heterotic string theory is
concerned, we can mention: the racetrack mechanism
\cite{Krasnikov:1987jj, Casas:1990qi}, the inclusion of
non-perturbative corrections to the Kaehler potential
\cite{Casas:1996zi, Binetruy:1996xja, Barreiro:1997rp}, the
inclusion of a downlifting sector \cite{Lowen:2008fm}. As far as moduli stabilization in heterotic-M-theory is concerned the reader is referred to \cite{Zanzi:2006xr, Correia:2007sv} and related articles.\\
Remarkably, in our model \cite{Zanzi:2010rs}, the E-frame dilaton parametrizes the amount of scale symmetry of the system. Therefore, the chameleonic behaviour of the field guarantees a scenario where scale invariance is abundantly broken locally (on short distance scales), while, on the contrary, scale invariance is almost restored on cosmological distances and, in particular, the E-frame CC is under control. Interestingly, this result is valid including all quantum contributions and without fine-tuning of the parameters. Happily, in the E-frame, the CC is under control and the dilaton is a chameleon, even if we consider in the string frame (S-frame) a very large vacuum energy with a stabilized dilaton: the necessary hierarchy between the S-frame CC and the E-frame one is produced after the conformal transformation (see also \cite{Zanzi:2012du}). For this reason we pointed out a non-equivalence of different conformal frames at the quantum level and we selected the E-frame as the physical one: in this model, the CC has a small positive value only in the E-frame (for a detailed discussion of the conformal transformations from the S-frame to the E-frame the reader is referred to \cite{Zanzi:2012du}). In the model of references \cite{Zanzi:2010rs, Zanzi:2012du, Zanzi:2012ha}, there are a number of consequences of this peculiar E-frame chameleonic scale invariance:\\ 1) the E-frame CC is under control.\\ 2) The concept of particle is re-examined \cite{Zanzi:2010rs, Zanzi:2012ha}: the string length is chameleonic (the string mass is an increasing function of the matter density). Local particles are the relevant degrees of freedom on short distances (i.e. locally) and they are small interacting strings. On the contrary, global (background) particles are the relevant degrees of freedom on very large cosmological distances and they are (almost non-interacting) cosmic strings \cite{Zanzi:2012ha}.\\ 3) E-frame matter fields are chameleonic \cite{Zanzi:2012ha}.

One relevant question is related to the theoretical origin of the model of references \cite{Zanzi:2010rs, Zanzi:2012du, Zanzi:2012ha}. This issue has been discussed, at least partially, in \cite{Zanzi:2012bf}: the model can be embedded, to some extent, in heterotic-M-theory and, under certain assumptions about the full M-theory action, the Casimir origin of the stabilizing potential for the S-frame dilaton has been pointed out in \cite{Zanzi:2012bf}. Moreover, the similarity of the dark matter lagrangian with a Ginzburg-Landau lagrangian for extradimensional neutrino condensation has been discussed in \cite{Zanzi:2012bf}. 
 
It seems worthwhile summarizing some open problems of our scenario for the CC problem.\\
A) A more detailed description of the QG aspects of the model would be welcome. After all, in \cite{Zanzi:2010rs} gravity is described through a metric and, hence, the reader might think that our description of gravity is semi-classical at best. Needless to say, it would be better to identify the fundamental pillars of our QG theory and a top-down approach would be rewarding.\\
B) It would be interesting to investigate the peculiar features of a theory respecting the fundamental principles of point (A).\\
C) The careful reader may be worried by the presence of multiple quantizations in \cite{Zanzi:2010rs} and this issue should be further discussed. Particular attention should be dedicated to potential connections with the non-equivalence of different conformal frames at the quantum level.\\
D) The renormalizability of our QG theory should be carefully discussed. This issue is particularly relevant in connection to the CC problem.\\
E) In \cite{Zanzi:2010rs} a crucial step of our procedure was a conformal transformation from the string frame to the Einstein one. It would be rewarding to understand whether this transformation is linked to some particular event of the cosmological evolution. Indeed, in this case, the conformal transformation would be easily included in the model.\\ 
F) It would be interesting to understand whether our chameleonic model predicts a peculiar pattern of soft terms. The breakdown of supersymmetry (SUSY) should be analyzed carefully.\\
G) The connection between our model and string theory should be further investigated. Particular attention should be dedicated to the fact that in the S-frame of \cite{Zanzi:2010rs} we are considering a strongly coupled theory and, therefore, the lagrangian must be exact (i.e. it must include all quantum loops).

In this paper we are going to analyze points A and B. {\it The remaining points C-G will not be discussed in this paper and they will be simply assumed to be solved.} Here is a summary of our results:\\
A) We are going to further analyze the theoretical grounds of the model of reference \cite{Zanzi:2010rs, Zanzi:2012du, Zanzi:2012ha, Zanzi:2012bf} and we are going to focus our attention on the chameleonic behaviour of matter fields. In particular, we will identify some requirements that produce as a final outcome a density-dependent mass of matter fields in a top-down approach. During this analysis, we will formulate a chameleonic equivalence postulate (CEP) as one of those requirements. In this way, the model of reference \cite{Zanzi:2010rs} is just one particular example of a set of models, induced by the CEP, where the mass of matter particles is density-dependent. Remarkably, the CEP is a microscopic counterpart of Einstein's Equivalence Principle and the gravitational aspects of the CEP are a consequence of the chameleonic nature of our model (in this way, the gravitational aspects of the CEP should not be considered as a postulate but as a principle). A new ''chameleonic'' description of gravity is obtained: quantum gravitation is equivalent to a conformal anomaly in our model. We suggest to exploit the CEP as a guideline towards QG and we establish a ''dictionary'' connecting the classical with the quantum regime.\\
B) We will point out that chameleon fields provide elements which are useful to understand the wave function collapse (for a review paper on the wave-function collapse the reader is referred to \cite{Bassi:2012bg}). The collapse is induced by the chameleonic nature of the theory and, as we will see, it is related to a shift of the ground state of the theory. Interestingly, in our model, the collapse of the wave-function is a QG effect. \\

As far as the organization of this article is concerned, in section 2 we discuss the CEP; section 3 discusses the connection between chameleon theories and quantum-mechanical wave-function collapse: in particular we will consider the model of reference \cite{Zanzi:2010rs} as an example of one model satisfying our CEP and in this framework we will analyze the collapse of the wave-function. Section 4 describes some gravitational aspects of the CEP. Some concluding remarks are discussed in section 5. The appendix summarizes a few elements about Bogoliubov transformations: these comments are useful to understand the projection operator relevant for the collapse of the wave-function.

\setcounter{equation}{0}
\section{Chameleonic Equivalence Postulate}
\label{CEP}

In the standard chameleonic literature, the Planck mass is typically constant and the chameleonic action is written in the E-frame. Matter fields are conformally coupled to the chameleon and the conformal factor guarantees the required competition between the scalar potential and the matter branch. For example, a typical choice that can be found in the literature is the exponential function $\rho_m e^{\beta \phi/M_p}$, where the matter density includes the mass $m_0$ of the matter field and $\beta$ parametrizes the coupling strength. Basically the exponential coupling with the scalar field can be interpreted as a mass-varying term for matter particles.

Now, let us suppose we are given a stringy model matching the following requirements: \\R1) the gravity part of the action can be summarized through a single scalar-tensor theory where a dilaton field is present (let us call it $\phi$ in the S-frame and $\sigma$ in the E-frame). \\R2) The E-frame dilaton $\sigma$ is chameleonic and it controls the strength of the couplings\footnote{This means, in particular, that the saturation mechanism for the couplings discussed in reference \cite{Veneziano:2001ah} is not part of our approach (at least at this stage).}.\\R3) Chameleonic Equivalence Postulate - CEP:
{\it for each pair of vacua V1 and V2 allowed by the theory there is a conformal transformation that connects them and such that the mass of matter fields $m_0,_{V1}$ (i.e. $m_0$ evaluated in V1) is mapped to $m_0,_{V2}$ (i.e. $m_0$ evaluated in V2). When a conformal transformation connects two vacua with a different amount of conformal symmetry, an additional term (in the form of a conformal anomaly) must be included in the field equations and this additional term is equivalent to the gravitational field.}

In order to render the physical content of the CEP easier to understand, we will proceed stepwise touching upon the  requirements mentioned above. Point (1) is very natural in a stringy model. Basically it is useful because we are simply getting rid of more complicated situations like multi-scalar-tensor theories. Point (2) is crucial and, in particular, it links the energy scale in the E-frame to the expectation value of the E-frame dilaton. From reference \cite{Zanzi:2012du} we infer that point (3) guarantees that the physical renormalized Planck mass is $\sigma$-dependent. Remarkably, even if we choose V1 and V2 in the same frame we can make a triangle V1-V3-V2 (where V3 belongs to a different frame with respect to V1 and V2) by composing two different conformal transformations. The first one connects V1 to V3 while the second one connects V3 to V2. The total transformation connects V1 to V2 in a single conformal frame. Since the mass of matter fields should be properly renormalized, $\sigma$-dependent couplings and a $\sigma$-dependent Planck mass imply, through renormalization, $\sigma$-dependent masses of matter fields. Therefore, a $\sigma$-dependence of the E-frame masses is naturally expected in a theory satisfying our three conditions. The chameleonic (the mass is an increasing function of the matter density) or anti-chameleonic (the mass is a decreasing function of the matter density) behaviour of matter particles is model-dependent and must be checked separately. In general, in a model where $\phi$ is not stabilized, a $\phi$-dependent renormalized Planck mass is expected: the renormalized Planck mass becomes a function of two variables ($\phi$ and $\sigma$).

For a discussion of the connection between CEP and QG, the reader is referred to section \ref{cepqg}.

In this paper we will be particularly interested in the lagrangian of \cite{Zanzi:2010rs} satisfying our requirements (R1-R3). In other words we write the string frame lagrangian as
\beq {\cal L}={\cal L}_{SI} + {\cal L}_{SB}, \label{Ltotale}\eeq where the
scale-invariant Lagrangian is given by:

\begin{equation}
{\cal L}_{\rm SI}=\sqrt{-g}\left( \half \xi\phi^2 R -
    \half\epsilon g^{\mu\nu}\partial_{\mu}\phi\partial_{\nu}\phi -\half g^{\mu\nu}\partial_\mu\Phi \partial_\nu\Phi
    - \frac{1}{4} f \phi^2\Phi^2 - \frac{\lambda_{\Phi}}{4!} \Phi^4
    \right).
\label{bsl1-96}
\end{equation}
$\Phi$ is a scalar field representative of matter fields,
$\epsilon=-1$, $\left( 6+\epsilon\xi^{-1} \right)\equiv
\zeta^{-2}\simeq 1$, $f<0$ and $\lambda_{\Phi}>0$.
One may write also terms like $\phi^3 \Phi$, $\phi \Phi^3$ and
$\phi^4$ which are multiplied by dimensionless couplings. However
we will not include these terms in the lagrangian. Happily, a $\phi^4$ term does not clash with the solution to the CC problem, because the renormalized Planck mass in the IR region is an exponentially decreasing function of $\sigma$ (see also \cite{Zanzi:2012du}).

To proceed further, let us discuss the symmetry breaking Lagrangian
${\cal L_{SB}}$, which is supposed to contain scale-non-invariant terms,
in particular, a stabilizing (stringy) potential for $\phi$ in the
S-frame. For this reason we write: \beq {\cal L}_{\rm
SB}=-\sqrt{-g} (a \phi^2 + b + c \frac{1}{\phi^2}). \label{SB}
\eeq

Happily, it is possible to satisfy the field equations with
constant values of the fields $\phi$ and $\Phi$ through a proper
choice (but not fine-tuned) values of the parameters
$a, b, c$, maintaining $f<0$ and $\lambda_{\Phi}>0$. This lagrangian will be further discussed in the remaining part of this article.

Interestingly, our CEP resembles the Equivalence Principle of quantum mechanics \cite{Faraggi:1997bd, Faraggi:1997yj, Faraggi:1998pc, Faraggi:1998pd}. Let us add some comments. Recently, the Quantum Stationary Hamilton-Jacobi equation (QSHJE) has been obtained in a top-down approach from the Equivalence Postulate/Principle (EP) of quantum mechanics. Remarkably, the QSHJE is a non-linear equation (like the chameleonic equations) and, once it is linearized, it produces the Schr$\ddot{o}$dinger equation. The potential connections between our CEP and the analysis of \cite{Faraggi:1998pd} should be investigated.

\setcounter{equation}{0}
\section{Chameleonic wave function collapse}
\label{debroglie}

Let us start from \ref{Ltotale}. For a discussion of the parameters of the model, the reader is referred to \cite{Zanzi:2012du}.
It seems worthwhile pointing out that a better fit with the Casmir energy of references \cite{Zanzi:2006xr, Zanzi:2012bf} can be obtained with a potential of the form ($A, B$ and $C$ are constants):
\bea
V_{SB}=\frac{A}{\phi^2}+B+\frac{C}{\phi},
\label{newcasimir}
\eea 
which, after dilaton stabilization, corresponds to a constant once again.

When we perform a conformal transformation to the E-frame (for a detailed discussion see \cite{Zanzi:2012du}), we can rewrite the lagrangian \reflef{bsl1-96}) as
\begin{equation}
{\cal L}_{*}=\sqrt{-g_*}\left( \frac{1}{2} R_* -
    \half g^{\mu\nu}_*\partial_{\mu}\sigma\partial_{\nu}\sigma + {\cal L}_{* matter}
    \right),
\label{eframe}
\end{equation}
where ${\cal L}_{* matter}$ turns out to be
\begin{equation}
{\cal L}_{* matter}= -
    \half g^{\mu\nu}_* D_{\mu}\Phi_* D_{\nu} \Phi_* - e^{2 \frac{d-2}{d-1} \zeta
    \sigma} (\xi^{-1} \frac{f}{4} M_p^2 \Phi_*^2+ \frac{\lambda_{\Phi}}{4!} \Phi_*^4)
\label{lmatter}
\end{equation}
and $D_{\mu}=\partial_{\mu}+ \zeta \partial_{\mu} \sigma$.

The chameleonic competition can be obtained in various ways. Let us summarize them explicitly.
\begin{itemize}
\item We can exploit a conformal-anomaly-induced interaction vertex between dilaton and matter (see \cite{Zanzi:2010rs} and related references). The coupling is basically $\rho_{M} \times \sigma$, where the matter density in \cite{Zanzi:2010rs} satisfies $\rho_{M} \propto e^{-4\zeta\sigma}$. The careful reader may be worried by this exponential damping of the matter density because, even if we manage to construct one ground state, this is not enough to render operative the chameleon mechanism. Indeed, whatever will be the matter density we choose, it must be possible to construct a corresponding ground state (for the ''table of this room'', for the ''air of this room''...). On the other hand, we cannot change the parameters of the model, because this modifies the lagrangian and it is our intention to write the lagrangian once and for all. Hence, an exponential damping of the matter density is, at first sight, a phenomenological problem of \cite{Zanzi:2010rs}. Happily, however, as we will show below, it is possible to construct a chameleonic minimum exploiting gauge fields and in this paper we will parametrize the matter density following \cite{Zanzi:2010rs}. It seems worthwhile pointing out that, once the parameters of the model are fixed, the matter density is determined by $\sigma$. A shift in the matter density is related to a shift of the dilatonic mass: if we increase $\rho_M$, the amount of scale symmetry is smaller, therefore, the mass-protection mechanism induced by the symmetry is less effective than before and $\sigma$ becomes heavier. As already mentioned in \cite{Zanzi:2010rs}, the dilaton is a chameleon in this model and this remains true even in the absence of a competition between the bare potential and the matter density: $\sigma$ is coupled to the trace of the energy-momentum tensor and its mass is an increasing function of the matter density because of symmetry reasons. Now we discuss the chameleonic competition obtained through the gauge fields. 
\item Let us start considering photons and let us consider the variation with respect to $\sigma$ of the relevant part of our E-frame lagrangian, namely, a dilatonic potential term $V(\sigma)$ and the $F^2$-term of photons. We can write: \beq \delta
S=\int d^4
x\sqrt{-g_*}\{-V,_{\sigma}(\sigma)\delta \sigma-\frac{1}{16
\pi} \frac{\partial \alpha_g^{-1}}{\partial \sigma} F_{\mu
\nu}F^{\mu \nu} \delta \sigma \} \nnb\\
=\int d^4
x\sqrt{-g_*}\{-V,_{\sigma}(\sigma)\delta\sigma-\frac{1}{16
\pi} \frac{\partial \alpha_g^{-1}}{\partial \sigma}
2(B^2-E^2)\delta\sigma \}
\nnb\\
\eeq where we introduced explicitly
the electric and magnetic fields through the formula \beq F_{\mu
\nu} F^{\mu \nu}=2 (B^2-E^2). \eeq We define\footnote{Do not confuse $\rho_{\gamma} \propto B^2-E^2$ with the electromagnetic energy density (proportional to the {\it sum} $B^2 +E^2$).} \beq \rho_{\gamma}=\frac{B^2-E^2}{8 \pi}.\eeq If we
require $\delta S=0$ we thus have the following term in the equation of motion for $\sigma$:
\begin{equation}
-V,_\sigma(\sigma)+
\frac{1}{\alpha^2(\sigma)}\frac{\partial \alpha}{\partial
\sigma}
\rho_{\gamma}.
\label{eqdilgenerale}
\end{equation}

Therefore, as far as the contribution of radiation to the effective dilatonic potential is concerned, the correct chameleonic competition between the two branches of the curve is present, granted that the electromagnetic coupling $\alpha_g(\sigma)$ is a decreasing function of $\sigma$. This is a natural requirement in our model and it is part of the requirement R2. A similar conclusion can be obtained for the other gauge fields and, in particular, for gluons. Therefore, inside standard matter, the presence of the chameleonic minimum is guaranteed by the competition with the $F^2$-term.
\end{itemize}

  The chameleonic behaviour of matter fields \cite{Zanzi:2012ha} is useful to justify the ansatz about the collapse of the wave-function in (non-relativistic) quantum mechanics. We will focus our attention on electrons and we will discuss separately\\
1) the role of non-linearities in a standard diffraction experiment with electrons;\\
2) the role of non-linearities in a Stern-Gerlach experiment;\\
3) the wave-function collapse in both experiments just mentioned above.

\subsection{Breakdown of the superposition principle in a electronic diffraction experiment}
\label{diff}

Let us consider a standard diffraction experiment with electrons where we produce a diffraction pattern on a screen and let us describe the results of this experiment through non-relativistic quantum mechanics. It is common knowledge that the position of one electron is well-defined {\it a posteriori}, namely, after the interaction electron-screen has taken place. Needless to say, this fact is compatible with the standard postulate of (non-relativistic) quantum mechanics regarding the collapse of the wavefunction of the electron (which is part of the so-called {\it Copenhagen interpretation} of the quantum formalism, see for example \cite{Ghirardi:1997xx}). To the best of our knowledge this postulate is still an open problem (at least to a certain extent) for the scientific community. Remarkably, in the non-relativistic quantum theory the collapse is supposed to take place {\it instantaneously}.\\
Now we come back to our model for a chameleonic dilaton. 

Remarkably, in our proposal, the collapse of the wave-function is related to a vacuum shift (i.e. to an energy density shift) in the chameleonic theory. This point we mentioned last needs to be further elaborated. \\It is well-known that in non-relativistic quantum mechanics the ket  $\mid\alpha >$ which represents a physical state can be written as a sum (i.e. {\it a superposition}) of position eigenvectors, namely (for the 1-dimensional case):
\bea
\mid \alpha>= \int dx \mid x> <x \mid \alpha>,
\label{superp}
\eea
where $<x \mid \alpha>$ is the wave-function of the system and $\mid <x\mid\alpha>\mid^2 dx$ is the probability that the particle is detected in a small interval $dx$ around $x$.  Needless to say, the superposition principle plays a crucial role when we write $\mid \alpha>$ as a sum of kets and this principle rests on the linearity of the wave equation. Let us now move to quantum field theory (QFT). Interestingly, in QFT the expectation value of a scalar matter field $\Phi_*$ (squared) is related to the particle number density. Therefore, in our diffraction experiment with electrons, the number density of the electrons on the screen can be parametrized, on the one hand, with the wave-function squared in the non-relativistic formalism, on the other hand, with the (expectation value of the) matter field squared in the relativistic formalism. Hence, we suggest in this way a connection between the non-relativistic wave-function and the relativistic quantum matter field\footnote{The reader may be interested in a historical introduction to QFT that can be found, for example, in \cite{Weinberg:libro}.} (this connection will be further discussed in paragraph \ref{single}). The linearity of the theory can be checked, on the one hand, in a non-relativistic formalism, by looking at the linear nature of the wave-equation and, on the other hand, in a relativistic formalism, by looking at the linear nature of the field equation. The relevant question is therefore: \\{\it what can we say about the linearity of our theory inside and outside the screen?} Once a certain matter density is fixed,  the chameleonic vacuum is well-defined and we can linearize the dilatonic theory around this vacuum: no matter whether we consider the behaviour of the dilaton inside or outside the screen, we reasonably assume that the theory can be well-approximated by linear equations obtained through harmonic approximations around the correct chameleonic vacuum (the ''inside one'' or the ''outside one''). Hence, the next question is:\\ {\it what can we say about the linear nature of our theory during a transition between two different chameleonic vacua (namely from the ''outside vacuum'' to the ''inside vacuum'')?} In this last case, strong non-linearities are expected in the field equations (the harmonic approximation is not valid anymore) and the superposition principle is broken for a short (but non-vanishing) interval of time. We infer that the superposition \ref{superp} cannot be valid anymore during the entrance of the electron in the screen (i.e. when the position measurement takes place) and, therefore, one single position eigenvector might be selected among the set \{x\} which we considered in the integration (i.e. the wave-function might collapse). This chameleon-induced breakdown of the superposition principle corresponds to a matter-density shift (i.e. to the entrance of the electron-string in the screen).

\subsection{Breakdown of the superposition principle in a Stern-Gerlach experiment}
  
 Naturally the postulate regarding the collapse of the wave-function must be applied to {\it all} measurement of (non-relativistic) quantum mechanics. In other words, whatever will be the hermitian operator whose spectrum we are interested in, the effect of the measurement must be described by this ansatz. Therefore, the diffraction example mentioned above is not general enough: it is very easy to imagine different quantum measurement where the collapse of the wave-function takes place in an environment with small matter density, for example a Stern-Gerlach (SG) experiment. 
 
 Let us discuss, therefore, a SG experiment with, for example, electrons. When the electrons are outside the magnet we have $\rho_{\gamma}=0$. When the electrons enter inside the magnet, a non-vanishing $\rho_{\gamma}$ is present, hence, the chameleonic competition between the E-frame dilatonic potential and $\frac{\rho_{\gamma}}{\alpha_g}$ increases the mass of the dilaton and reduces the amount of scale-symmetry. In other words, when we perform the (spin) measurement, we electromagnetically interact with the electrons, therefore, a shift in $\rho_{\gamma}$ is obtained. In our chameleonic model it is precisely this $\rho_{\gamma}$-shift that induces (through the dilaton) a breakdown of the superposition principle in a small transition region (whose size is comparable with the string length) between the two chameleonic vacua 1) the $B=0$ vacuum and 2) the $B \neq 0$ vacuum. We now apply to the spinorial case the same discussion developed for the position eigenvectors. In both chameleonic vacua (the $B=0$ one and the $B\neq 0$ one) we reasonably assume that it is possible to linearize the theory, but strong non-linearities are expected in the transition region. Before the entrance in the magnet, we choose the ket which represents the electron as a superposition of spin-up and spin-down kets, namely, with obvious notations:
\bea
\mid \alpha> = \frac{\mid + > + \mid ->}{\sqrt{2}},
\label{input}
\eea the theory is linear and the chameleonic vacuum is the $B=0$ one. Inside the magnet, the chameleonic vacuum is the $B \neq 0$ one and the theory is linear once again. However, when the electron-string enters the magnet (i.e. when the spin measurement takes place), we have a transition between the two chameleonic vacua, hence, non-linearities are important and, therefore, the superposition principle, which plays a crucial role in the sum \ref{input} between up and down kets, is not valid anymore. The theory might choose for this reason one single ket (i.e. the spinorial wave-function might collapse). In the next section we will further discuss this ''selection mechanism''. 

\subsection{Wave function collapse}

Let us come back to the collapse. In our analysis, we managed to show that, given an initial ket 
for the system (written as a superposition of various kets), during the transition between two different chameleonic vacua, a breakdown of the superposition principle is expected when non-linearities are properly taken into account. However, if our intention is to justify the collapse through chameleons, this is not enough. For example, one crucial question that is still waiting for an answer is: why the breakdown of the superposition principle should imply the selection of one single ket? 

Basically, in our approach we are discussing connections between different formalisms:\\ 1) the classical formalism;\\ 2) the quantum non-relativistic formalism; \\3) the QFT formalism.\\4) String theory.\\ 

We will proceed stepwise discussing (1) the origin of the projection operator responsible for the collapse of the wave function and (2) our two previous experimental configurations, namely the SG and the diffraction experiment with electrons.

\subsubsection{Projection operator} 

We point out that, during the jump between an initial and a final configuration $\sigma_i \Rightarrow \sigma_f$ (or $\Phi^*_i \Rightarrow \Phi^*_f$), the vacuum energy is not conserved. The vacuum energy of the system is chameleonic and, consequently, a non-unitary evolution might be present in the chameleonic jump between two vacua\footnote{As far as a connection between unitarity and energy conservation is concerned, the reader is referred to the analysis of the scattering theory discussed in \cite{Weinberg:libro}.}. To proceed further, we explore the possibility that this potentially non-unitary evolution is connected with a non-unitary projection operator. As already discussed in the literature \cite{Giulini:1996nw}, if we introduce in a model a set of projection operators, we are splitting the Hilbert space in a set of subspaces, typically characterized by {\it unitarily inequivalent} representations of a certain algebra of observables. Therefore, a connection between chameleon fields and superselection rules\footnote{A superselection rule guarantees, by definition, that the system occurs in states, for example, $\mid \psi_1 >$ and $\mid \psi_2 >$ but never in a superposition  $a\mid \psi_1 > +  b\mid \psi_2 >$.} might take shape.

The careful reader may be interested in a more detailed and formal discussion of the projection operator in our model. To illustrate this point we exploit Bogoliubov transformations \cite{Bogoliubov:1958xx}. Let us start considering vanishing number density in the chameleonic model: no particles. If our intention is to create particles, we must apply creation operators to the vacuum. Needless to say, the creation operators are related to the quantum fields. Once a non-negligible number of matter particles have been created\footnote{This discussion can be repeated also for the magnetic field in the SG experiment, even if in that case the chameleon mechanism is not due to the matter density. The presence of the magnetic field gives a non-vanishing local vacuum energy and, once again, the state of minimum energy is not the state annihilated by all the annihilation operators.}, the number density (hence the local vacuum energy) is non-vanishing and the state of minimum energy is {\it not} annihilated by all the annihilation operators. In other words, this chameleonic model is telling us that ''in this room'' we live in a state which is a minimum of the chameleonic effective potential, but the matter density is non-vanishing. At this stage, our system resembles the free electron gas model (FEGM) discussed in appendix, granted that we compare (1) the electron of the FEGM with the matter particle of the chameleonic model and (2) the Fermi energy of the FEGM with the non-vanishing local vacuum energy of the chameleonic model. Now the matter density (hence the local vacuum energy) is non-vanishing anymore and, therefore, if we want to create more particles we have to exploit creation operators which are {\it different} with respect to the creation operators we started with: the creation/annihilation operators we started with do not respect the Fock condition when the local vacuum energy is non vanishing. The two sets of operators are related to each other by a Bogoliubov transformation (see the appendix). As already mentioned above, creation/annihilation operators of matter particles are related to matter fields. We infer that the two sets of quantum fields (pre and post measurement) are related to each other by a Bogoliubov transformation. Happily, the general Bogoliubov transformation is not unitary in the infinite volume limit (see \cite{Strocchi:1985cf}): this non-unitary transformation is, in our proposal, the projection operator necessary for the collapse of the wave-function in our model and, remarkably, it is connected to a variation of the vacuum energy.

To proceed further, we discuss an alternative derivation of the projection operator exploiting projective representations of the conformal group. Indeed, the shift in the vacuum energy is related to the conformal factor and therefore it must be possible to summarize the presence of a superselection rule through the conformal factor. This idea is fully compatible with the CEP because the jump between the two vacua (before/after measurement) is performed through the composition of two different conformal transformations: (1) from the vacuum before the measurement to the string frame vacuum and (2) from the string frame vacuum to the vacuum after the measurement.

\subsubsection{Projective representations of the conformal group}

When a quantum measurement is performed, the dilaton is stabilized through the chameleon mechanism. As already discussed in \cite{Zanzi:2010rs, Fujii:2003pa}, our dilaton $\sigma$ is a pseudo-Nambu-Goldstone boson of broken scale invariance. The mass of $\sigma$ is related to an explicit breakdown of scale invariance. The dilaton can get a mass in two ways in our model: 1) through the conformal anomaly induced $\sigma\Phi_*^2$ term (as already mentioned above, even if $\sigma \Phi_*^2$ is exponentially damped, a shift in the matter density is related to a shift in the amount of scale symmetry and, hence, of the dilatonic mass); 2) through the competition with the $F^2$ term of gauge fields. Since both terms produce a chameleonic mass for the dilaton (i.e. explicit breaking of scale invariance), we summarize both terms through the presence of a central charge in the algebra of the conformal group and this is one of the possible ways projective representations may present themselves (the other way is topological).
Indeed, it is common knowledge that the appearance of central charges is the counterpart for the algebra of the presence of phases in a projective representation of a group, namely, the elements $T, \bar{T}, etc.$ of a symmetry group are represented on physical Hilbert space by unitary operators $U(T), U(\bar{T}),etc.$ which satisfy the composition rule
\bea
U(T)U(\bar{T})=e^{i\phi(T,\bar{T})} U(T,\bar{T}),
\eea 
where $\phi$ is a real phase (for an introduction to projective representations the reader is referred to \cite{Weinberg:libro}). The double conformal transformation mentioned in the previous paragraph (from the vacuum before the measurement to the string frame and from the string frame to the final vacuum after the measurement) corresponds to $U(T)U(\bar{T})$.

The relevant result, for our purposes, is the following. If it is possible to prepare the system in a state represented by a linear combination, then the phases $\phi(T, \bar{T})$ cannot depend on which of these classes of states the operators  act upon (for a proof of this result the reader is referred again to \cite{Weinberg:libro}, chapter 2). Equivalently, if the phases $\phi$ may depend on the state, then there must be a superselection rule. 
In our model, the central charge is a function of the fields and therefore it is related to the microscopic quantum state of the local vacuum which, as already mentioned above, contains also the particles of ''this room'' where the measurement is performed. Therefore, the phases of the projective representation of the conformal group are functions of the various quantum observables (mass, spin, angular momentum...). In other words, the various quantum numbers we can measure are encoded into the corresponding fields ($\Phi_*$, $A_{\mu}$...) and the interaction of those fields with $\sigma$ will produce a dependence of the phase $\phi(\sigma)$ on the microscopic quantum state of the theory (which includes the system we measure). In this way {\it a superselection rule is obtained when the conformal anomaly is non-negligible} (i.e. when the dilaton is stabilized) and, hence, the wave-function collapses through the chameleon mechanism. In our proposal, when the measurement is performed, a certain amount of energy is transferred to the system and, therefore, a field stabilization is expected through the chameleon mechanism (after the measurement).

The careful reader may be worried by these ideas because it is well-known that it may or it may not be possible to prepare a system in a linear combination of states, but one cannot settle the question by reference to a symmetry principle, because whatever will be the symmetry group we use, there will be always another group with identical consequences except for the absence of superselection rules (see \cite{Weinberg:libro}, page 90). However, to the best of our knowledge, this remark we mentioned last does not apply with conformal anomalies and, consequently, we still summarize the projection operator through the presence of a non-trivial phase in a projective representation of the conformal group. 

To proceed further, we will discuss the SG and the diffraction experiment.

\subsubsection{Stern-Gerlach and diffraction experiment}
\label{single}

\begin{itemize}
\item
{\it Non-degenerate case}\\
As already mentioned above, we connect relativistic fields to non-relativistic wave-functions. The purpose of this section is to render this connection more precise. The connection is based on the number density of particles $n_p$. On the one hand, in non-relativistic quantum mechanics $n_p$ is parametrized by the absolute value squared of the wave-function and, on the other hand, in a QFT language, it is parametrized by the expectation value (squared) of the field (at least for a scalar field).
Remarkably, expectation values of quantum operators (quantum fields) are spatial integrals weighted by the integration volume (i.e. spatial averages). 
Let us now consider the SG configuration and let us suppose we measure the z-component of the spin operator.
One problem must be faced with our chameleon-induced collapse.
If we justify the collapse exploiting the chameleon mechanism, when the particles escape from the magnetic field, the chameleon mechanism is not operative, the dilaton is not stabilized and the solution to the chameleonic field equation is expected to be the same we had before the measurement (namely, before the entrance into the magnet). On the other hand, it is well known that, after a SG measurement is performed, the particles are split into up and down particles (no linear combinations of up and down are allowed) and this result is valid even if we start with a linear combination of up and down kets before the measurement. Hence, if we prepare the system of electrons in a linear combination of up and down kets and we let them enter into the magnetic field, the question is: why the particles escaping from the magnet (in the absence of magnetic field) are not described by the same ket they had before the entrance into the magnetic field (namely by the linear combination of up and down kets)? This question is relevant if our intention is to support the connection between quantum fields and non relativistic wave function mentioned above. 
The answer is that the number density of electrons outside the magnetic field doesn't know anything about the spin state. We can have up or down or a linear combination of up and down kets, but the number density is the same. Hence, the expectation value of the field is (roughly) the same we had before the measurement. 

Remarkably, whenever we interpret an expectation value (squared) as a number density, we are implicitly considering a macroscopic average of the field because, in order to talk about number density, we must consider a large number of particles and, hence, macroscopic length scales. The situation is reminiscent of thermodynamics (to a certain extent), where a single value of a macroscopic averaged quantity (like the temperature for example) is compatible with a large number of microscopic quantum states. These microscopic quantum states are encoded in our model into the quantum fields (it appears in a QFT language with the expansion of a field in terms of creation and annihilation operators), while the averaged macroscopic quantities are related to the number density (to the expectation values of the fields). 

These comments are supporting the idea that the superselection rules, already obtained through projective representations of the conformal group, should be related to the number density. This comment must be further elaborated. 
 Let us suppose that the up and down states have different energy: this seems to be very reasonable in the SG configuration.  In our SG example, inside the magnet, formally we write
\bea
H \mid +> = E_+ \mid +>\\
H \mid -> = E_- \mid ->,
\eea
where $E_+ \neq E_-$. When the field is sitting at the minimum (inside the magnet), there is no dynamical evolution. Therefore, we suggest to connect this stabilized field configuration where the expectation value is constant with a stationary state of the non-relativistic formalism. In our SG example, the system is forced to choose either the up or the down ket, because a linear combination of the two states would not be stationary inside the magnet: it would correspond to a non-stationary wave-function inside the magnet, it would give a non-trivial time evolution to the absolute value squared of the wave function, this would force the macroscopic number density to change in time and this would clash with the constant expectation value of the stabilized field. 
In other words, a linear combination of kets is forbidden after the measurement: it is not compatible with the environment which stabilizes the fields and, hence, it is forbidden because it is not a stationary (stabilized) state.  

\item
{\it Degenerate case}\\
Typically, when the quantum system is symmetric, degenerate states occur. Let us now consider the diffraction experiment with electrons and let us suppose we are equipped with an axially-symmetric experimental set-up. In other words, we consider an initial electron beam (i.e. a plane wave) and we study the diffraction of the electrons through a circular hole. The expected diffraction pattern will respect the circular symmetry (i.e. it corresponds to a set of degenerate states), but the single electron on the screen will {\it break} the symmetry. {\it Where does this symmetry breaking come from?} First of all, let us come back to our chameleon fields. We point out that the lagrangian is rotationally symmetric. What can we say about the vacuum? If we manage to show that the vacuum is not invariant under rotations, we can claim that the rotational symmetry is spontaneously broken. This is exactly the path that we are going to follow here: spontaneous breakdown of rotational symmetry will be useful to justify the diffraction pattern on the screen.\\
Naturally the matter density defines the vacuum through the chameleon mechanism. Some comments are necessary to proceed further.\\
1) Even if we fix the value of the matter density, the length scale compatible with that particular value of $\rho_m$ is not unique. In particular, the matter density of the screen is the correct matter density on macroscopic length scales (e.g. the length of the screen), but also on microscopic length scales (e.g. the atomic radius).\\
2) As already mentioned above (see \cite{Zanzi:2012ha}), the string length of the electron is a decreasing function of the matter density. Before the measurement is performed, if the matter density is small enough, the string length of the electron in the beam is larger than the atomic radius and the relevant matter density during the measurement is roughly 1 $g/cm^3$. After the entrance in the screen, the string becomes shorter, but its length must be always compatible with a matter density of (roughly) 1 $g/cm^3$.  \\
3) Whatever will be the Feynman diagram we consider, after the various quantization steps in the E-frame, the UV cut-off of the theory is provided by the string mass \cite{Zanzi:2012ha}.\\   
Now, to the point. What is the UV cut-off of the theory inside the screen (after the measurement)? In other words, what is the shortest possible length scale compatible with a matter density of, basically, $1$ $g/cm^3$? The answer is: the atomic radius. Remarkably, the string length inside the screen is comparable with the atomic radius which, needless to say, is comparable to the de Broglie wavelength of the atomic electrons.\footnote{An interesting line of development will try to interpret the probability waves of quantum mechanics as waves on quantum strings.} Therefore, the UV cut-off inside the screen is basically the inverse of the atomic radius.\\

On atomic length scales, namely on length scales similar to the UV cut-off of the theory, rotational symmetry is broken: this is the origin of the spontaneous breakdown of rotational symmetry inside the screen. To illustrate this point, let us connect the E-frame lagrangian of \cite{Zanzi:2010rs} to the Olive-Pospelov lagrangian \cite{Olive:2007aj}:

\begin{eqnarray}
S_{\phi} = \int d^4x \sqrt{-g} \Biggl\{- \frac{M_{\rm Pl}^2}{2} R + \frac{{M_*}^2}{2}
\partial^{\mu} \phi
\partial_{\mu} \phi  - V(\phi) \nonumber \\
-  \frac{B_{F}(\phi)}{4}F_{\mu\nu}F^{\mu\nu} + \sum_{j=n,p,e}
[\bar\psi_j iD\!\!\!\!/ \psi_j -
B_j(\phi)m_j\bar\psi_j\psi_j]\Biggr\}. \label{lagrangian}
\end{eqnarray}

In other words, we replace the anomaly-induced coupling of \cite{Zanzi:2010rs} between $\sigma$ and the matter density (where the matter fields were represented by a scalar field $\Phi_*$) with a coupling between $\sigma$ and a matter density represented by a spinor field (like in the Olive-Pospelov approach). To proceed further, we connect the spinorial matter fields to the wave-function following chapter 14 of \cite{Weinberg:libro} and we write the renormalized matter density as:
\bea
m_j \bar\psi_j(<\sigma>) \psi_j(<\sigma>) &\simeq& m_j <\Phi_N \mid \psi_j^{\dagger} \mid \Phi_0> <\Phi_0 \mid \psi_j \mid \Phi_N> + BR + ...\nn\\ &=& m_j u^{\dagger}_N u_N + BR +...= m_j \mid u_N \mid^2 + BR + ...,
\eea
where $\mid \Phi_0>$ is the vacuum, $\mid \Phi_N>$ is an atomic state vector, ''BR'' represents the contribution of the counterterm (the backreaction, see \cite{Zanzi:2010rs}) and the dots represent the terms of the form $m_j <\Phi_N \mid \psi_j^{\dagger} \mid \Phi_i> <\Phi_i \mid \psi_j \mid \Phi_N>$ ($\mid \Phi_i>$ is a set of orthonormal states that is not complete because the vacuum is missing). We infer that the matter density is related to the wave-function squared. Needless to say, the contribution of atomic electrons is not rotationally invariant and, therefore, the matter density shares the same property. Consequently, the amount of scale invariance is slightly modified whenever a rotation is performed: the final result is a non-symmetric renormalized local vacuum energy. We infer that the vacuum does not respect rotational symmetry inside the screen and, hence, rotational symmetry is spontaneously broken inside the screen. 

There are a number of consequences of this fact. A single point on the screen selects a direction in space. The connection between a non-relativistic formalism and the relativistic one pointed out above, suggests that the selection of one particular direction in this case, which takes place after the collapse, is analogous to the selection of one particular direction by a bended rod with circular cross section placed vertically on a table and pushed with a (strong enough) force over it (the classical example discussed in QFT textbooks when introducing spontaneous symmetry breaking, see for example \cite{Ryder:1985wq}).\\ 
\\
When a large number of electrons is considered, the symmetric pattern is restored and the same behaviour is shown when a rod is bended a very large number of times (the corresponding quantum vacuum is symmetric once again). \\
\end{itemize}

Interestingly, our approach is reminiscent of environment-induced decoherence \cite{Giulini:1996nw, Hepp:1972ii, Auletta, Schlosshauer:2003zy}. Another interesting line of development will try to clarify whether our loss of unitarity in the theory is restored at a more fundamental level and will study potential connections with the information problem in black-holes. As already mentioned above, a detailed phenomenological analysis is definitely required.

\section{CEP and quantum gravity}
\label{cepqg}

Let us discuss some gravitational aspects of the CEP.

The theory of Special Relativity (SR) is based on an invariance principle: the laws of Nature are invariant under Lorentz transformations. In other words, whatever will be the inertial frames we consider, we are free to perform a Lorentz transformation that connects two inertial systems and this transformation cannot change the equations of the theory. Needless to say, SR is not a theory of gravity. If our intention is to describe gravity in a relativistic way, we can consider GR and the Equivalence Principle (EP) plays a crucial role. The EP is telling us that inertia is equivalent to gravitation. It is common knowledge that whenever we perform a transformation that brings us from an inertial frame to a non-inertial one, additional terms will be present in the equations of the theory. A non-relativistic example is given by the Newton equations: in non-inertial frames the Newton equations acquire some additional terms (the inertial forces). This idea is valid
also in GR. When we perform a general coordinate transformation in GR, some ''additional terms'' (the metric and the connection) will be present in the equations of the theory and these terms are exploited by Einstein to describe the gravitational field in harmony with the EP. GR is a classical theory of gravity.

Let us now discuss the quantum regime and let us start with a question: how is it possible to describe a complete absence of gravity? Our GR-based intuition tells us that we must remove all the masses and all the energy sources (including vacuum energy!). The 4D cosmological (IR) vacuum discussed in this paper provides one ground state where gravity is basically absent. If our intention is to switch on a (small) gravitational field we can add a source in the form, for example, of a massive (or even massless) particle and the amount of conformal symmetry will be slightly reduced. For example, if we add an electron, the related contribution to the matter density will reduce the amount of conformal symmetry while, if we add a photon, the coupling to the $F^2$-term will reduce the amount of conformal symmetry. In this way, whenever we modify the gravitational field, we obtain a chameleonic shift of the ground state. In other words, in our model, the chameleon mechanism is telling us that the gravitational field is described by the conformal anomaly in harmony with the CEP. Our CEP connects different vacua where a different degree of conformal symmetry is present. We suggest to exploit the CEP as a guideline towards the QG regime. In particular, let us construct the following dictionary for QG:
\begin{itemize}
\item let us replace the inertial frame of Einstein's theories with a conformal ground state in our chameleonic model. Let us write the connection between the two models in this way: inertial frame $ \rightarrow$ conformal ground state. 
\item Non-inertial frame $\rightarrow$ non-conformal ground state.
\item General coordinate transformation $\rightarrow$ conformal transformation.
\item Metric and connection $\rightarrow$ conformal anomaly.
\end{itemize}

In this way, the ''dictionary'' mentioned above creates a connection between classical and QG: A) conformal transformations in QG are analogous to general coordinate transformations in classical gravity; B) the cocycle in QG is analogous to the metric and the connection in classical gravity. The CEP is the microscopic counterpart of the EP.\\
A few comments are in order.\\
1) Remarkably, our description of gravity through a conformal anomaly is telling us how to deal with vacuum energy in QG. A shift in the vacuum energy can be summarized by a conformal anomaly through the chameleon mechanism.\\
2) Another question is related to the holographic description of the model. As already discussed by \cite{Dvali:2012en} in the framework of quantum N-portrait, whenever we consider AdS/CFT correspondence, the central charge of the CFT is the occupation number of AdS gravitons. Our model is developed from the standpoint of heterotic-M-theory and the 5D bulk is (almost) AdS. In other words, our CEP is giving us an ''additional term'' to the field equations, namely the cocycle (the conformal anomaly), and this term is related to the occupation number of gravitons in 5D. Once again, conformal anomaly is gravity.\\
3) As already discussed in \cite{Weinberg:1972gr}, the question ''why gravitation should obey the EP?'' does not find an answer inside GR. On the contrary, in our chameleonic model, we can consider the equivalence of gravitation and conformal anomaly as an outcome of our analysis. Therefore, in our model, (the gravitational aspects of) the CEP should not be considered as a postulate, but as a final result. A fast exponential damping of the matter density and the related problems in obtaining a minimum of the effective potential do not clash with our conclusions because we can exploit the shift in the amount of conformal symmetry: as already mentioned above, if we add one electron, the amount of conformal symmetry is slightly smaller and, therefore, the vacuum energy and the mass of $\sigma$ are (slightly) larger or, in other words, a chameleonic shift is obtained.\\
4) Let us come back for a moment to the wave function collapse. From the elements we gathered, we infer that the collapse of the wave-function is a QG effect.\\
5) Needless to say, scale invariance plays a key role in our model. The $\sigma$-dependence of the amount of symmetry is even more important: we have an abundantly broken symmetry in the UV region that becomes restored in the IR. Therefore, a natural question is: do we have similar theories in the literature? Interestingly, see for example \cite{Blaszczyk:2011ib} for a recent discussion, heterotic string compactifications can be conveniently described in the language of gauged linear sigma models (GLSMs). GLSMs are not conformal, because in two dimensions the gauge couplings and some kinetic terms have non-vanishing mass dimension, but it is believed that in the IR limit the theory flows to a conformal model. A promising line of development will investigate potential connections between our model and the papers \cite{Blaszczyk:2011ib, McOrist:2010ae, Adams:2012sh, Witten:1993yc, Nakayama:2004vk}. \\
6) Our description of QG resembles, to a certain extent, QCD. For example, we know that the mass of a baryon is related to $\Lambda_{QCD}$ (and hence to a conformal anomaly) and it is due, to a large extent, to the presence of strong interaction (the quark masses give a very small contribution). In other words, the conformal anomaly plays a crucial role also in the low-energy description of strong interaction. Moreover, in our model, the Planck mass is a condensation scale (see also \cite{Zanzi:2012bf}) and dilaton stabilization (fixing the value of the dilaton) is a gauge-fixing for our QG theory analogously to gauge-fixing in non-abelian gauge theories. Moreover, if we exploit S-duality in our scenario, our QG theory is expected to be free in the deep UV region. We will further analyze these issues in the future.\\

An interesting line of development will analyze the microscopic nature of backreaction (the counterterm in our model), but now we already know that backreaction is equivalent to a conformal anomaly and, in other words, to gravity, because backreaction changes the amount of conformal symmetry of the ground state. A good starting point for a more detailed and formal analysis is given by \cite{Maldacena:2003nj, Barvinsky:1994cg, Nakayama:2004vk}. In particular, in \cite{Maldacena:2003nj}, the cocycle $S(g_1)-S(g_2)$ is written as a Liouville action ($g_1$ and $g_2$ are two metrics related to each other by a conformal transformation). 

Remarkably, the vacuum energy in our model gives a repulsive contribution to the particles, because the vacuum energy is minimized for small number densities. This contribution might be relevant because we do not fine tune the scale of the potential to small values. An interesting line of development will further analyze this aspect of our chameleonic approach to QG. We will try to understand whether it is possible to reinterpret our gravitational theory through entropic arguments (a good starting point would be \cite{Verlinde:2010hp}). For example, we could try to study the free expansion of a gas of particles (or a gravitational collapse problem) taking into account also the chameleonic behaviour of the vacuum energy. A repulsive vacuum energy should be present also in the Sun and in the Earth and might give us some interesting phenomenological signature of chameleon fields.

\section{Conclusions and possible lines of development}
\label{conclusions}

In this paper we discussed these points:\\
A) We further analyzed the theoretical grounds of the model of reference \cite{Zanzi:2010rs, Zanzi:2012du, Zanzi:2012ha, Zanzi:2012bf} and we focused our attention on the chameleonic behaviour of matter fields. In particular, we identified some requirements that produce as a final outcome a density-dependent mass of matter fields in a top-down approach. During this analysis, we formulated a chameleonic equivalence postulate (CEP) as one of those requirements. In this way, the model of reference \cite{Zanzi:2010rs} is just one particular example of a set of models, induced by the CEP, where the mass of matter particles is density-dependent. Remarkably, the CEP is a microscopic counterpart of Einstein's Equivalence Principle and the gravitational aspects of the CEP are a consequence of the chameleonic nature of our model (in this way, the gravitational aspects of the CEP should not be considered as a postulate but as a principle). A new ''chameleonic'' description of gravity is obtained: quantum gravitation is equivalent to a conformal anomaly. We suggested to exploit the CEP as a guideline towards QG and we established a ''dictionary'' connecting the classical with the quantum regime.\\
B) We pointed out that chameleon fields provide elements which are useful to understand the wave function collapse (for a review paper on the wave-function collapse the reader is referred to \cite{Bassi:2012bg}). The collapse is induced by the chameleonic nature of the theory and it is related to a shift of the ground state of the theory. Interestingly, in our model, the collapse of the wave-function is a QG effect. \\

A detailed phenomenological analysis of the entire model is required to test these ideas. 

Some comments are in order.

1) One problem might arise if we apply the Veneziano's mechanism of reference \cite{Veneziano:2001ah, Zanzi:2012ha} and we saturate the fundamental couplings. It remains to be seen whether the many copies of the SM clash with the chameleon-induced collapse of the wave-function. Another comment is in order. As already mentioned above, in this model the E-frame dilaton must be large enough in order to generate reasonable values of matter density but, at the same time, $\sigma$ must be in the non-perturbative regime to make the Veneziano's mechanism operative. We should check that the chameleonic collapse of the wave function passes the dangerous phenomenological tests related to the variation of fundamental couplings. A detailed phenomenological analysis is necessary. At the moment, we simply point out that the mass of matter fields is planckian and, hence, there is a nice cancellation between $G_N$ and the Planck mass (squared) in the non-relativistic formula of the gravitational force between two matter particles.

2) As far as the chameleonic competition between the potential and the radiation term is concerned, we notice that $E^2$ and $B^2$ enter into the effective potential with opposite signs and, in particular, only the magnetic field corresponds to a stabilizing chameleonic contribution. This minus sign and its potential consequences should be further discussed.

Among the possible lines of development we can mention:\\
a) As far as the chameleon-induced collapse of the wave function is concerned, a detailed phenomenological analysis is definitely required to test these ideas. It would be interesting to understand whether the theory restores unitarity at a more fundamental level.  \\ 
b) The ansatz about the collapse of the wave function enters in every single measurement process of quantum mechanics, therefore, our chameleon-induced collapse together with the CEP can be considered as a step towards the construction of a {\it chameleonic quantum mechanics}. These issues should be investigated. For example, one possibility would be to reanalyze quantum neutron interferometry experiments in the presence of a magnetic field from the standpoint of this chameleonic model (for a non-chameleonic discussion of these experiments the reader is referred to \cite{Sakurai:1985ll} and related references). We will also try to establish deeper connections between our proposal and references \cite{Bassi:2012bg, Faraggi:1998pd, Ghirardi:1997xx}. Moreover, we will try to find an answer to this question: is it possible to introduce in the theory a chameleonic Planck constant $h=h(\sigma)$? \\
c) The chameleonic behaviour of matter particles should be further analyzed and particular attention should be dedicated to its potential phenomenological signatures.\\
d) Potential connections between chameleons and Black-Holes' physics should be analyzed starting from \cite{Giulini:1996nw, Bassi:2012bg, Hsu:2009ve, Hsu:2010sb, Susskind:2005js, Hepp:1972ii, Barrow:libro, Blanchard:libro, Dvali:2008sy, Dvali:2011aa}.\\
e) Potential connections between our model and references \cite{Hepp:1972ii, Faraggi:1998pd, Strocchi:1985cf, Auletta, Schlosshauer:2003zy, Giulini:1996nw, DeWitt:2003pm, Susskind:2005js, Barbon:2005jr, Barbon:2003aq, 'tHooft:2012xh, tHooft:2011aa, Parikh:1999mf} should be investigated.\\
f) Another interesting line of development will analyze potential connections between chameleon fields and fractals.\\
g) An interesting future research direction will try to clarify whether probability waves of quantum mechanics can be interpreted as waves on quantum strings.\\
h) It would be interesting to construct a stabilizing potential for the S-frame dilaton exploiting an extension of the Seiberg-Witten \cite{Seiberg:1994aj, Seiberg:1994rs} theory to D=5 N=2 Supergravity.  Happily, for this theory, gravitational quantum corrections have been analyzed in \cite{Antoniadis:1995vz, Antoniadis:1995jv, Antoniadis:1995ct} where a D=5 N=2 Seiberg-Witten theory coupled to gravity has been discussed in the framework of heterotic theory and therefore the resulting potential is exact (a condition which must be satisfied by our dilatonic potential in the S-frame, because we are in the strong coupling regime of string theory and hence perturbation theory cannot be exploited). However, particular attention is necessary with the contribution of massive string states. A promising line of development will further discuss the connection between our model and references \cite{Antoniadis:1995vz, Antoniadis:1995jv, Antoniadis:1995ct}.\\
i) A future research direction will search for potential connections between the three quantization steps of our proposal and references \cite{RoblesPerez:2012rj, Buonanno:1996um, Peleg:1993vw, McGuigan:1989tn}.\\
l) It would be interesting to understand whether chameleonic QG is able to tell us why Nature obeys the Principle of Inertia in classical physics. During this analysis, potential connections between our model and Mach's principle should be investigated.\\
m) It would be interesting to understand whether we can construct a chameleonic graviton. Perhaps reference \cite{Hamber:2013rb} might be useful.

\subsection*{Acknowledgements}

Special thanks are due to Dieter Luest, Pieralberto Marchetti, Antonio Masiero, Marco Matone, Marios Petropoulos, Massimo Pietroni and Augusto Sagnotti. They were a great source of inspiration during the development of this article.
 
I thank also Ignatios Antoniadis, Marco Bochicchio, Denis Comelli, Michele Redi, Mario Tonin and Roberto Volpato for useful comments and discussions.

\appendix
\section{Bogoliubov transformation and free electron gas}

In this appendix we touch upon some useful formulas regarding Bogoliubov transformations and their application to the free electron gas. For a more detailed discussion the reader is referred to \cite{Strocchi: 1985cf}.

Let us start considering a certain number $N$ of electrons in a cubic box of side $L$. Whatever will be the electron we choose, the wave vector ${\bf k}$ can take discrete values $k_i=(2\pi/L)n_i$, $n_i=0,1,2...$. The ground state of the system is the lowest energy state compatible with the Pauli principle. Remarkably, the Fermi energy $E_F$ is related to the number density $n=N/V$ and, in particular, we can write
\bea
E_F=\frac{\hbar ^2}{2m}(3\pi^2 n)^{2/3}.
\eea
In this system, the state of minimum energy is {\it not} the state annihilated by all the annihilation operators. In other words, the ground state (i.e. the state of minimum energy) does not satisfy the Fock condition with respect to the electron annihilation operators and, indeed, for $k<k_F$ we have:
\bea
a(k,s) \Psi_0\neq0, 
\eea
where $k$ is the momentum and $s$ is a spin variable.
This implies that $a$ and $a^*$ do not describe the elementary excitations of the free electron gas. Interestingly, it is possible to perform a Bogoliubov transformation in order to introduce new creation and annihilation operators such that the ground state $\Psi_0$ satisfies the Fock condition for the new operators.
In general a Bogoliubov transformation can be written as
\bea
A_k=u_k a_k-v_ka^*_{-k}\\
A^*_k= u_k^*a_k^*-v^*_ka_{-k}.
\eea
In this picture the excitation of an electron from the ground state ($k<k_F$) to a state with $k>k_F$ is described by the creation of (1) an excited state above the Fermi surface with energy $E(k)-E_F$ and (2) the creation of a hole (i.e. an unoccupied state inside the Fermi sphere) with energy $\mid E(k)-E_F \mid$. A general Bogoliubov transformation in the infinite volume limit is {\it non-unitary} (see for example \cite{Strocchi:1985cf}).

\addcontentsline{toc}{section}{{Bibliography}}
\providecommand{\href}[2]{#2}\begingroup\raggedright\endgroup

\end{document}